\documentclass{article}

\usepackage[utf8]{inputenc}
\usepackage{tikz}
\usetikzlibrary{decorations.pathreplacing}
\usepackage{MnSymbol}
\usepackage{paralist}
\usepackage[a4paper]{geometry}
\usepackage{amsmath}
\usepackage{amsthm}
\usepackage{mathrsfs}

\renewcommand{\phi}{\varphi}

\newcommand{\Aut}[1]{\mathscr{#1}}

\newcommand{\Dual}[1]{\text{compl}(#1)}

\newcommand{\SCC}[1]{\text{SCC}(#1)}

\newcommand{\VarsTL}{\mathcal V}

\newcommand{\Next}{{\medcircle}}
\newcommand{\TLNext}{{\mathsf{X}}}

\newcommand{\Eventually}{\mathsf{F}}

\newcommand{\Always}{\mathsf{G}}

\newcommand{\Until}[2]{#1 \mathsf{U} #2}

\newcommand{\Release}[2]{#1 \mathsf{R} #2}

\newcommand{\TransConds}[2]{\text{TC}(#1,#2)}

\newcommand{\NuTL}{\nu\text{TL}}
\newcommand{\MuTL}{\mu\text{TL}}

\newcommand{\Pair}[2]{\langle #1, #2\rangle}
\newcommand{\Norm}[1]{\text{norm}(#1)}

\newcommand{\Neg}[1]{\text{neg}(#1)}

\title{Backward deterministic and weak alternating $\omega$-automata\footnote{This work was supported by DFG.}}
\author{Sebastian Preugschat, Thomas Wilke\\
  Kiel University, Germany\\
  \texttt{thomas.wilke@email.uni-kiel.de}}

\theoremstyle{plain}
\usepackage{amsthm}
\newtheorem{theorem}{Theorem}
\newtheorem{corollary}{Corollary}

\newtheorem{fact}[theorem]{Fact}

\theoremstyle{remark}
\newtheorem{example}{Example}

\begin{document}

\maketitle

\begin{abstract}
  We present a direct transformation of weak alternating $\omega$-automata into equivalent backward deterministic $\omega$-automata and show (1) how it can be used to obtain a transformation of non-deterministic Büchi automata into equivalent backward deterministic automata and (2) that it yields optimal equivalent backward deterministic automata when applied to linear-time temporal logic formulas. (1) uses the alternation-free fragment of the linear-time $\mu$-calculus as an intermediate step; (2) is based on the straightforward translation of linear-time temporal logic into weak alternating $\omega$-automata.
\end{abstract}

\section{Introduction}

It is only natural to read $\omega$-words from left to right: they have a definite start, but no end, so one reads one letter after the other, starting with the letter in the first position. This is probably why we typically envision an $\omega$-automaton as a device that when running over an $\omega$-word starts by consuming the letter in the first position, then goes over to the letter in the second position, then to the letter in the third position, and so on. We call this the \emph{forward approach}. Almost all of $\omega$-automata theory is based on the forward approach, in particular, there is a huge body of work on the determinization of $\omega$-automata, where---if one wanted to be precise---determinization means the process of constructing automata which are deterministic when following the forward approach.

There is essentially one fundamental result in the theory of $\omega$-automata with regard to the \emph{backward approach}, where automata start in the infinite and run until they reach the beginning of the word: Olivier Carton and Max Michel \cite{carton-michel-2003} proved that every regular $\omega$-language is recognized by a backward deterministic Büchi automaton. In other words, in the backward approach---unlike in the forward approach---all types of acceptance conditions classically considered (Büchi, generalized Büchi, parity, Rabin, Streett, Muller) give rise to the same class of $\omega$-languages and this class is the same as the class of $\omega$-languages recognized by non-deterministic automata (expressive completeness). Technically, the main contribution of Carton and Michel are two different constructions that turn a given non-deterministic Büchi automaton into an equivalent backward deterministic generalized transition Büchi automaton (which they show can be transformed into an equivalent backward deterministic Büchi automaton). The main contribution of the present paper is a direct transformation of a given \emph{forward} weak alternating $\omega$-automaton into an equivalent backward deterministic generalized transition Büchi automaton.

Weak alternating automata can be viewed as alternation-free formulas of the modal $\mu$-calculus, and vice versa. This was first demonstrated by André Arnold and Damian Niwiński for sets of infinite trees~\cite{arnold-niwinski-1992} and carries over to $\omega$-words directly. For $\omega$-words, it is moreover true that weak alternating automata and the alternation-free fragment of the modal $\mu$-calculus---often referred to as $\NuTL$ \cite{banieqbal-barringer-1987, vardi-1988} or $\MuTL$ \cite{lange-2005}---are expressively complete in the sense that they both describe exactly the class of all regular $\omega$-languages.
(Note that the alternation hierarchy of $\NuTL$ collapses on the second level \cite{arnold-niwinski-1990}, whereas on trees it is strict \cite{bradfield-1998, lenzi-1996}.) Expressive completeness follows from \cite{arnold-niwinski-1992} and work by Kupferman and Vardi on weak alternating $\omega$-automata \cite{kupferman-vardi-2001} and was also proved by Kaivola \cite{kaivola-1995}.

Based on all this we describe a new way to convert a given non-deterministic Büchi automaton into an equivalent backward deterministic Büchi automaton: we go from a Büchi automaton to a $\NuTL$-formula, then from $\NuTL$ to weak alternating automata (immediate), and finally apply our construction. In some sense, we break down the construction from~\cite{carton-michel-2003} into simpler constructions at the expense of complexity. The automata we construct are somewhat larger than the ones constructed by Carton and Michel.

Formulas of linear-time temporal logic (LTL) are typically translated into non-deterministic Büchi automata \cite{vardi-wolper-sistla-1983}; a standard translation will actually produce a backward deterministic automaton \cite{preugschat-wilke-2013}. There is, however, also a straightforward way to translate an LTL formula into a weak alternating automaton \cite{gastin-oddoux-2001, muller-saoudi-schupp-1988}. So our construction can serve to obtain a backward deterministic automaton for a given LTL formula: simply apply the construction to the weak alternating automaton obtained from a given LTL formula. We show that the automaton thus obtained has the same size as the ``standard automaton''.

% The remainder of the paper is structured as follows. In Section~2, the main result is presented. Section~3 covers the applications of the main result as described above.

\section{From weak alternating to backward deterministic automata}

In this section, we present our main result, a transformation from weak alternating to backward deterministic $\omega$-automata. We begin with basic definitions and results we draw on. 

\subsection{Weak alternating $\omega$-automata}
\label{sec:weak-altern-omega}

There are different ways of formalizing weak alternating automata; the variant used in this paper works with transition conditions rather than a partition of the state space into existential and universal states.

Given a set $Q$ of states and an alphabet $A$, the \emph{transition  conditions} over $Q$ and $A$ are formulas built from 
\begin{itemize}
\item $B$, for $B \subseteq A$, and
\item $\Next q$, for $q \in Q$,
\end{itemize}
using the boolean connectives $\vee$ and $\wedge$. The set of all these conditions is denoted by $\TransConds Q A$.

A \emph{weak alternating automaton} \cite{muller-saoudi-schupp-1986} over an alphabet $A$ is given by
\begin{itemize}
\item a finite set $Q$ of \emph{states,}
\item a \emph{transition function} $\delta \colon Q \to \TransConds Q A$, and
\item a partition of the state set $Q$ into a set $R$ of recurring and a set $N$ of non-recurring states. (More formally, a pair $\langle  R, N \rangle$ is given such that $R
  \cup N = Q$ and $R \cap N = \emptyset$ hold. The elements of the first component are called \emph{recurring states} and the elements of the second component are called \emph{non-recurring states.})
\end{itemize}
In addition, there is a requirement on recurring and non-recurring states with regard to the \emph{transition graph}. This graph is the directed graph with vertex set $Q$ and an edge from $q$ to $q'$ if $\Next q'$ occurs in $\delta(q)$.

The requirement is that either $S \cap Q \subseteq R$ or $S \cap Q \subseteq N$ holds for every strongly connected component $S$ of the transition graph, that is, all states of any strongly connected component (SCC) must be recurring or else non-recurring.

A \emph{very weak alternating automaton} is one where the SCC's of the transition graph are singleton sets (which immediately implies that very weak alternating automata are weak alternating automata).

In general, runs of alternating automata are labeled trees satisfying certain conditions, but since the recurrence condition we work with can be viewed as a Büchi or parity condition, it is sufficient to consider graphs, as described in what follows, see~\cite{kupferman-vardi-2001}.

A \emph{run graph} of an automaton as described above on a word $u$ is a directed graph where the vertices are pairs of the form $\langle i, \tau\rangle$ with $i \in \omega$ and $\tau$ is a state or a subformula (including the formula itself) of any of the transition conditions $\delta(q)$.

The following conditions must be satisfied for every vertex $\langle i, \tau \rangle$:
\begin{itemize}
\item If $\tau \in Q$, then $\langle  i, \tau\rangle$ has exactly one outgoing edge and this leads to $\langle  i, \delta(\tau) \rangle$.
\item If $\tau = \Next q$, then $\langle  i, \tau \rangle$ has exactly one outgoing edge and this leads to $\langle  i+1, q \rangle$.
\item If $\tau = \tau_0 \vee \tau_1$, then $\langle  i, \tau\rangle$ has outgoing edges only to $\langle i, \tau_0\rangle$ and $\langle i, \tau_1\rangle$ and at least one such edge exists.
\item If $\tau = \tau_0 \wedge \tau_1$, then $\langle  i, \tau\rangle$ has outgoing edges exactly to $\langle i, \tau_0\rangle$ and $\langle i, \tau_1\rangle$.
\end{itemize}
In addition, vertices $\langle i, \tau\rangle$ with $\tau = B$ and $u(i) \notin B$ must not exist.

A good way to envision these graphs is to imagine the vertices arranged in levels numbered 0, 1, 2, \dots, where on level $i$ the vertices of the form $\Pair i q$ are grouped together and between level $i$ and level $i+1$ the vertices $\langle  i, \tau \rangle$ with $\tau$ being a transition condition are located. Then the edges only go from vertices on level $i$ through \emph{intermediate vertices} to vertices on level~$i+1$.

For every infinite path through the run graph, the set $Q'$ of states occurring infinitely often in it is a subset of an SCC of the transition graph, which means $Q' \subseteq R$ or $Q' \subseteq N$. If $Q' \subseteq R$, the path is said to be \emph{final}. The run graph is said to be \emph{final} if all infinite paths through it are final. 

For every $i$, the suffix $u[i,\omega)$ is said to be \emph{accepted} from $q$ in a run graph if the graph is final and $\Pair i q$ is a vertex of it; it is \emph{accepted} from $q$ by the automaton if there exists a final run graph that accepts it from~$q$.

In  Subsection~\ref{sec:construction} we use a result on complementing alternating automata. To state it, we first define, for every transition condition, the \emph{complementary transition condition} by an appropriate set of equations:
\begin{align}
  \Dual B & = A \setminus B\\
  \Dual {\Next q} & = \Next q\\
  \Dual {\tau_0 \vee \tau_1} & = \Dual{\tau_0} \wedge \Dual{q_1}\\
  \Dual {\tau_0 \wedge \tau_1} & = \Dual{q_0} \vee \Dual{q_1}
\end{align}
The automaton \emph{complementary} to a given automaton $\Aut A$, denoted
$\Dual{\Aut A}$, is determined as follows:
\begin{itemize}
\item It has the same set of states as $\Aut A$.
\item Its transition function, denoted $\Dual \delta$, is defined by
\begin{align}
  \Dual \delta(q) = \Dual{\delta(q)}, \text{for every $q \in Q$.}
\end{align}
\item The sets of recurring and non-recurring states are exchanged.
\end{itemize}

The fact we need is the following one.

\begin{fact}[complementation of alternating automata, \cite{muller-schupp-1987}]
  \label{thm:complementation}
  Let $\Aut A$ be a weak alternating automaton over some alphabet $A$ and $q$ some state of it. For every $u \in A^\omega$ and $i < \omega$, the suffix $u[i,\omega)$ is accepted by $\Aut A$ from state~$q$ if, and only if, $u[i,\omega)$ is not accepted by $\Dual{\Aut A}$ from state~$q$.
\end{fact}

\subsection{Backward deterministic $\omega$-automata}

In general, a \emph{backward deterministic automaton} is given by 
\begin{itemize}
\item a finite set $Q$ of states,
\item a transition function $\rho \colon A \times Q \to Q$, and 
\item a \emph{recurrence condition} $\Omega$, which can be any acceptance condition such as a Büchi or a Muller condition, state-based or transition-based (see below).
\end{itemize}
A \emph{run} of such an automaton on a word $u \in A^\omega$ is a word $r \in Q^\omega$ such that $r(i) = \rho(u(i), r(i+1))$ holds true for every $i \in \omega$. A run $r$ is \emph{final} if it satisfies the recurrence condition. For instance, if $\Omega \subseteq Q$ is a (state-based) Büchi condition, then $r$ is final if there exist infinitely many $i$ such that $r(i) \in \Omega$. For a backward deterministic automaton, it is required that for every $u \in A^\omega$ there is exactly one final run!

In the following fundamental theorem, automata are viewed as defining sets of $\omega$-words: an automaton is augmented by a set $I \subseteq Q$ of initial states and then \emph{defines} the set of all $u \in A^\omega$ where $u$ is accepted from some state $q \in I$ (for weak alternating automata) or where $r(0) \in I$ is true for the unique final run of the automaton on~$u$ (for backward deterministic automata).

% \begin{fact}[completeness of Büchi condition, \cite{carton-michel-2003}]
%   \begin{enumerate}
%   \item For every Büchi automaton with $n$ states there exists an
%     equivalent backward deterministic generalized transition Büchi automaton with at most $(3n)^n$ states.
%   \item For every backward deterministic generalized transition Büchi automaton with $(3n)^n$ states and an acceptence condition of cardinality $2n$ there exists a backward deterministic Büchi automaton with at most $(12n)^n$ states.
%   \end{enumerate}
% \end{fact}

\begin{fact}[completeness, \cite{carton-michel-2003}]
  For every Büchi automaton with $n$ states there exists an equivalent backward deterministic generalized transition Büchi automaton with at most $(3n)^n$ states and an equivalent backward deterministic Büchi automaton with at most $(12n)^n$ states.
\end{fact}

\subsection{Main result}

To describe our main result we view weak alternating automata and backward deterministic automata as devices defining functions rather than languages. This is more general and gives a clearer result.

Let $\Aut A$ be a weak alternating automaton over some alphabet $A$ and with state set $Q$. The function \emph{computed} by $\Aut A$, denoted $f_{\Aut A}$, is the function $A^\omega \to (2^Q)^\omega$ where $f(u)(i)$ is the set of all $q$ such that the suffix $u[i,\omega)$ is accepted from $q$.

Let $\Aut A$ be a backward deterministic automaton, $B$ some alphabet,  and $\lambda \colon Q \to B$ an \emph{output} function. The function \emph{computed} by $\Aut A$ with respect to $\lambda$, denoted $f_{\Aut A, \lambda}$, is the function $A^\omega \to B^\omega$ defined by $f(u) = \lambda(r(0)) \lambda(r(1)) \dots$ where $r$ is the unique final run of $\Aut A$ on $u$. 

\theoremstyle{plain}
\newtheorem*{mainresult}{Main Theorem}
\begin{mainresult}
  For every weak alternating automaton $\Aut A$ there exists a backward deterministic automaton $\Aut B$ and an output function $\lambda$ for $\Aut B$ such that the function computed by $\Aut A$ is the same as the function computed by $\Aut B$ with respect to $\lambda$, that is, $f _ {\Aut A} = f _ {\Aut B, \lambda}$. The automaton $\Aut B$ has the following properties.
  \begin{enumerate}
  \item Let $S_0, \dots, S_{k-1}$ be an enumeration of all SCC's of the transition graph of $\Aut A$ and $m_i = S_i \cap Q$ for $i<k$. Then the number of states of $\Aut B$ is at most $\prod_{i<k} (m_i + 1)^{m_i}$, in particular, $(n+1)^n$ is an upper bound for the number of states of $\Aut B$ when $n$ is the number of states of $\Aut A$.
  \item The automaton $\Aut B$ has a generalized transition Büchi condition with as many Büchi sets as $\Aut A$ has states.
  \item The automaton $\Aut B$ has at most $2^n$ states when $\Aut A$ is a very weak alternating automaton with $n$ states.
  \end{enumerate}
\end{mainresult}

\subsection{The construction}
\label{sec:construction}

In this section, we describe how the automaton $\Aut B$ from the Main Theorem is constructed. We proceed in three steps, starting with a basic automaton and refining it twice. The first refinement takes care of problems with non-recurring states; the second refinement takes care of problems with recurring states.

In general, given an $\omega$-word $u$, the automaton $\Aut B$ tries to determine, for each position~$i$, the states from which the suffix $u[i,\omega)$ is accepted by $\Aut A$. This is why we model a state of~$\Aut B$ as assigning an appropriate value $v_q$ to each state $q \in Q$, indicating the ``degree'' of accepting (or non-accepting) $u[i,\omega)$ from~$q$. Formally, a state is a family $\{v_q\}_{q \in Q}$. 

\subsubsection{Basic approach} 

% \enlargethispage{\baselineskip}

In the basic approach the ``variables'' $v_1$ are boolean variables, more precisely, $v_q \in \{1, \infty\}$, where $1$ stands for true and $\infty$ for false. 

The rules from the definition of a run graph as given in Subsection~\ref{sec:weak-altern-omega} can be turned immediately into rules describing how the variables must be updated when a letter is read. To describe this precisely, we assume a state $\{v_q\}_{q \in Q}$ is given. We want to define $\rho(a, \{v_q\}_{q \in Q})$, which we write as $\{v'_q\}_{q \in Q}$. 

With every transition condition $\tau$, we associate a corresponding expression $e[\tau]$, defined by the following set of equations:
\begin{align}
  e[B] & = \text{if $a \in B$ then $1$ else $\infty$} \\
  e[\Next q] & = v_q\\
  e[\tau_0 \vee \tau_1] & = \min\{e[\tau_0], e[\tau_1]\}\\
  e[\tau_0 \wedge \tau_1] & = \max\{e[\tau_0], e[\tau_1]\}
\end{align}
Then $v'_q$ is obtained by evaluating $e[\delta(q)]$.

\paragraph*{Shortcomings of the basic approach.} 

The transition function defined above has several shortcomings, as described in what follows.

Consider a weak alternating automaton $\Aut A$ with a single state $q$ which is non-recurring and where $\delta$ is defined by $\delta(q) = \Next q$. First, there are two runs of $\Aut B$, namely one in which $v_q = 1$ holds all the time and another one where $v_1=\infty$ holds all the time. Second, in the run graph of $\Aut A$ corresponding to the first run of $\Aut B$, the state $q$ occurs infinitely often on a path, which means the run graph is not final and must be ruled out.

One way to approach these two problems is to introduce an appropriate recurrence condition in~$\Aut B$, corresponding to the partition of the state space of $\Aut A$ in recurring and non-recurring states. In fact, when, for instance, we require that $v_1=\infty$ holds infinitely often, the problems just mentioned disappear. But if there is more than just one state in an SCC of $\Aut A$, a recurrence condition does no longer help by itself, as explained in what follows. 

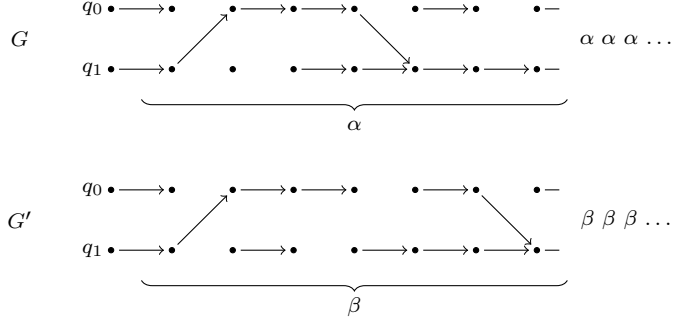
\begin{figure}[t]
  \centering
  \footnotesize
  \begin{tikzpicture}[every node/.style={circle,fill,draw},outer sep=2pt,inner sep=0.7pt,scale=0.8]
    \begin{scope}[shift={(0,0.5)}]
    \node[fill=none,draw=none] at (-0.5,3) {$G$};
    \node[fill=none,draw=none] at (0.7,3.5) {$q_0$};
    \node[fill=none,draw=none] at (0.7,2.5) {$q_1$};
    \foreach \x in {1,2,3,4,5,6,7,8} \node[draw,fill] (d\x) at (\x, 3.5) {};
    \foreach \x in {1,2,3,4,5,6,7,8} \node[draw,fill] (c\x) at (\x, 2.5) {};
    \draw[->] (c2) -- (d3);
    \draw[->] (d5) -- (c6);
    \foreach \x/\y in {1/2, 3/4, 4/5, 6/7} \draw[->] (d\x) -- (d\y);
    \foreach \x/\y in {1/2, 4/5, 5/6, 6/7, 7/8} \draw[->] (c\x) -- (c\y);
    \node[fill=none,draw=none] (c9) at (8.5,2.5) {};
    \node[fill=none,draw=none] (d9) at (8.5,3.5) {};
    \node[fill=none,draw=none] at (9.5,3) {$\alpha$ $\alpha$ $\alpha$ \dots};
    \draw (c8) -- (c9);
    \draw (d8) -- (d9);
    \draw[decorate,decoration={brace,mirror,amplitude=4pt}] (1.5,2) -- (8.5,2) node[midway,yshift=-10pt,fill=none,draw=none] {$\alpha$};
 \end{scope}
    %%%
    \node[fill=none,draw=none] at (-0.5,0.5) {$G'$};
    \node[fill=none,draw=none] at (0.7,1) {$q_0$};
    \node[fill=none,draw=none] at (0.7,0) {$q_1$};
    \foreach \x in {1,2,3,4,5,6,7,8} \node[draw,fill] (b\x) at (\x, 1) {};
    \foreach \x in {1,2,3,4,5,6,7,8} \node[draw,fill] (a\x) at (\x, 0) {};
    \draw[->] (a2) -- (b3);
    \draw[->] (b7) -- (a8);
    \foreach \x/\y in {1/2, 3/4, 4/5, 6/7} \draw[->] (b\x) -- (b\y);
    \foreach \x/\y in {1/2, 3/4, 5/6, 6/7, 7/8} \draw[->] (a\x) -- (a\y);
    \node[fill=none,draw=none] (a9) at (8.5,0) {};
    \node[fill=none,draw=none] (b9) at (8.5,1) {};
    \node[fill=none,draw=none] at (9.5,0.5) {$\beta$ $\beta$ $\beta$ \dots};
    \draw (a8) -- (a9);
    \draw (b8) -- (b9);
    \draw[decorate,decoration={brace,mirror,amplitude=4pt}] (1.5,-0.5) -- (8.5,-0.5) node[midway,yshift=-10pt,fill=none,draw=none] {$\beta$};    
  \end{tikzpicture}
  \caption{Illustration of the shortcomings of the basic idea, see Example~\ref{ex:shortcomings}}
  \label{fig:basic}
\end{figure}

\begin{example} 
  \label{ex:shortcomings}
  Imagine a weak alternating automaton $\Aut A$ which has one SCC $S$ consisting of two non-recurring states, $q_0$ and $q_1$. There could be two words, $u$ and $u'$, such that the graphs~$G$ and~$G'$ depicted in Figure~\ref{fig:basic} are the run graphs of $\Aut A$ on $u$ and $u'$, respectively. (Note that $G$ and $G'$ are two infinite graphs which are the results of repeating a finite graph infinitely often.) 

  The graph $G$ is not final, while $G'$ is. The two graphs can, however, not be distinguished by any recurrence condition. The two states occur infinitely often in $G$ and in $G'$, so there is no state-based recurrence condition that works. Moreover, for $G$ and $G'$ the sets of all subgraphs induced by two consecutive levels and occurring infinitely often coincide. This means that no recurrence condition based on transitions works either.\qed
\end{example}

\subsubsection{Non-recurring states---first improvement}

In Example~\ref{ex:shortcomings} we have seen that a run of $\Aut B$ may correspond to a run graph in $\Aut A$ which is not final, because there may exist paths through this run graph with infinitely many non-recurring states. The idea for solving this problem is explained in what follows, where we ignore recurring states for the moment; they are dealt with in a dual fashion in the second improvement below.

Assume $v$ is a vertex of a final run graph $G$ of $\Aut A$, say $v = \langle i, q\rangle$, and $q$ is non-recurring. Then the subgraph $H$ of $G$ consisting of the vertices reachable from $v$ and with second component in $\SCC q$ is finite (because we are dealing with weak alternating automata). Let $v'$ be a successor of $v$ in $G$. Either $v'$ does not belong to~$H$ anymore or we can consider the subgraph $H'$ of $H$ consisting of the vertices reachable from $v'$. The graph $H'$ is a proper subgraph of $H$, that is, its size is smaller. Now we can pass from $H'$ to $H''$ in the same way  we passed from $H$ to $H'$ (or stop because no vertices are left) arriving at an even smaller subgraph. This process must eventually terminate because $H$ was finite and the size of the subgraphs decreases in every step.

This is why in our construction, we measure the ``size'' of the subgraphs $H$, $H'$, $H''$, \dots\ and make sure that the size decreases along any path until another SCC or the end of the path is reached. Here, ``size'' does not simply mean number of vertices, because this could be unbounded. The notion of size we use is coarser. So, to be precise, the size may also stay the same from one vertex of a path to the next, but we enforce it to to decrease eventually, which is obviously enough for the construction to be correct.

What size exactly means can be deduced from the proof of correctness of our construction in Subsection~\ref{sec:correctness}. Here, we only describe the construction.

Let $S$ be an SCC of the transition graph and assume $S \subseteq N$. The variables $v_q$, for $q \in S$, now have values in $\{1, \dots, |S|, \infty\}$. The values $v'_q$ are determined in two steps: 
\begin{enumerate}
\item Values $\tilde v_q$ are determined where also $0$ is allowed as a value. 
\item These values are ``lifted'' so as to obtain values in the above range.
\end{enumerate}
The details are spelled out further below, after some more terminology. 

We use a normalization function when passing from one SCC to another one, in order to make sure that the values do not become to large. This function is denoted $\text{norm}$ and defined by $\Norm \infty = \infty$ and $\Norm i = 0$ for $i \in \omega$. We also use $q \leftrightsquigarrow q'$ to denote that $q$ and $q'$ belong to the same SCC of the transition graph.

To determine $\tilde v_q$ we adapt the expressions from above:
\begin{align}
  \label{eq:50}
  e'[B] & = \text{if $a \in B$ then $0$ else $\infty$} \\
  \label{eq:51}
  e'[\Next q'] & = \text{if $q \leftrightsquigarrow q'$ then $v_{q'}$ else $\Norm{v_{q'}}$}\\
  e'[\tau_0 \vee \tau_1] & = \min\{e'[\tau_0], e'[\tau_1]\}\\
  e'[\tau_0 \wedge \tau_1] & = \max\{e'[\tau_0], e'[\tau_1]\}
\end{align}
We let $\tilde v_q$ be the value obtained by evaluating $e'[\delta(q)]$.

The above equations are the same as the ones used in the basic approach, but there are two differences. First, for atomic formulas of the form $B$, the value $0$ is used to model that a formula is true immediately, see \eqref{eq:50}. Second, the normalization is incorporated to make sure the automaton starts all over again when passing from SCC to another one, see \eqref{eq:51}. 

The \emph{lifting} works as follows. Let $m$ be the minimal value $\geq 0$ such that $m$ does not occur as a value of one of the $\tilde v_q$'s for $q \in S$. If $m = 0$, then $v'_q = \tilde v_q$ for all $q \in Q'$. Otherwise, for every $q \in Q'$,
\begin{align}
  \label{eq:9}
  v'_q = \text{if $\tilde v_q > m$ then $\tilde v_q$ else $\tilde v_q + 1$.}
\end{align}
We say that $m$ is the \emph{critical value} of the transition with respect to the SCC~$S$.

Observe, firstly, that the lifting ensures that all values $v'_q$ are greater than~$0$ and that the order of the $v_q$'s is the same as the order of the $\tilde v_q$'s. Observe, secondly, that no finite value greater than $|S|$ can occur by the above definition---the worst case is when $\infty$ does not occur but every value between $1$ and~$|S|$. Observe, thirdly, that the else branch in \eqref{eq:9} is the one (and only one) place where finite values greater than $1$ are generated.

\begin{figure}[t]
  \centering
  \footnotesize
  \begin{tikzpicture}[every node/.style={fill,draw},outer sep=2pt,inner sep=0.7pt,scale=1]
    \node[fill=none,draw=none] at (0.7,1) {$q_0$};
    \node[fill=none,draw=none] at (0.7,0) {$q_1$};
    \foreach \x/\y in {1/1,2/1,3/2,4/2,5/1,6/2,7/2,8/1} \node[draw=none,fill=none] (b\x) at (\x, 1) {$\y$};
    \foreach \x/\y in {1/2,2/2,3/1,4/1,5/2,6/2,7/2,8/2} \node[draw=none,fill=none] (a\x) at (\x, 0) {$\y$};
    \draw[->] (a2) -- (b3);
    \draw[->] (b7) -- (a8);
    \foreach \x/\y in {1/2, 3/4, 4/5, 6/7} \draw[->] (b\x) -- (b\y);
    \foreach \x/\y in {1/2, 3/4, 5/6, 6/7, 7/8} \draw[->] (a\x) -- (a\y);
    \node[fill=none,draw=none] (a9) at (8.5,0) {};
    \node[fill=none,draw=none] (b9) at (8.5,1) {};
    \node[fill=none,draw=none] at (9.5,0.5) {$\beta$ $\beta$ $\beta$ \dots};
    \draw (a8) -- (a9);
    \draw (b8) -- (b9);
    \draw[decorate,decoration={brace,mirror,amplitude=4pt}] (1.5,-0.5) -- (8.5,-0.5) node[midway,yshift=-10pt,fill=none,draw=none] {$\beta$};    
%    \foreach \x in {1,2,3,4,5,6,7,8} \node[draw=none,fill=none] at (\x, 0) {};
  \end{tikzpicture}
  \caption{Illustration of the improved approach, see Example~\ref{ex:improved}}
  \label{fig:improved}
\end{figure}
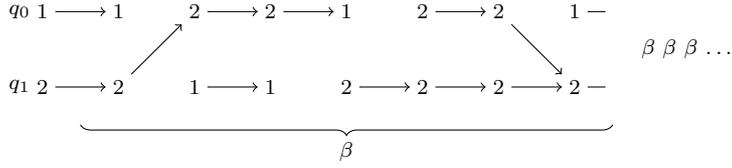

\begin{example}[Example~\ref{ex:shortcomings} continued] 
  \label{ex:improved}
  In Figure~\ref{fig:improved}, the graph $G'$ from Figure~\ref{fig:basic} is decorated according to the improved transition function.\qed
\end{example}

The recurrence condition we use is a generalized transition Büchi condition that makes sure every finite value $v_q$ originates from another SCC or a transition condition of the form $B$ for some $B \subseteq A$. Using the above terminology it makes sure that the graph $H$ is indeed finite. Technically, this means the values of the variables decrease down to~$1$ and then ``disappear'' at some point.

For every $i$ with $0 < i \leq |S|$, there is a transition Büchi set $B_{S,i}$ containing all transitions satisfying at least one of the following two conditions.
\begin{itemize}
\item The number $i$ is the critical value of the transition with respect to $S$.
\item There is no $q \in S$ such that $v'_q \geq i$.
\end{itemize}

\begin{example}[Example~\ref{ex:shortcomings} continued] 
  \label{ex:improved2}
  In Figure~\ref{fig:improved}, the first transition does not belong to any Büchi set, the second belongs to $B_{\{q_0,q_1\},1}$, the third does not belong to any, the fourth belongs to $B_{\{q_0,q_1\},2}$, \dots\ . This indicates a final run graph.\qed
\end{example}

\subsubsection*{Recurring states---second improvement}

With recurring states, there are similar problems as with non-recurring states.

\begin{example}[Example~\ref{ex:shortcomings} reused] 
  Assume that in Example~\ref{ex:shortcomings} the states $q_0$ and $q_1$ would be recurring. In the run graph $G$ the automaton could guess that all states on the infinite path are states from which the respective suffixes are not accepted, which is not true. But no recurrence condition, neither state- nor transition-based, could rule this out.\qed
\end{example}

To deal with these problems, we use the same approach as with non-recurring states but interpret the values of the variables complementary. That is, $\infty$ means the respective suffix is accepted, a finite value means it is not. To implement this, we make use of duality as stated in Fact~\ref{thm:complementation}.

Technically, this means that we have two sets of defining rules, which are dual to each other, depending on whether the respective state is non-recurring or recurring. 

The equations determining the value of the variable $\tilde v_q$ for a non-recurring state $q$ are:
\begin{align}
  \label{eq:30}
  e''[B] & = \text{if $a \in B$ then $0$ else $\infty$} \\
  e''[\Next q'] & = \text{if $q \leftrightsquigarrow q'$ then $v_{q'}$ else (if $q' \in N$ then $\Norm{v_{q'}}$ else $\Neg{\Norm{v_{q'}}}$)}\\
  e''[\tau_0 \vee \tau_1] & = \min\{e''[\tau_0], e''[\tau_1]\}\\
  \label{eq:33}
  e''[\tau_0 \wedge \tau_1] & = \max\{e''[\tau_0], e''[\tau_1]\} 
\end{align}
The ones for a recurring state $q$ are completely dual:
\begin{align}
  e''[B] & = \text{if $a \in B$ then $\infty$ else $0$} \\
  e''[\Next q'] & = \text{if $q \leftrightsquigarrow q'$ then $v_{q'}$ else (if $q' \in R$ then $\Norm{v_{q'}}$ else $\Neg{\Norm{v_{q'}}}$)}\\
  e''[\tau_0 \vee \tau_1] & = \max\{e''[\tau_0], e''[\tau_1]\}\\
  e''[\tau_0 \wedge \tau_1] & = \min\{e''[\tau_0], e''[\tau_1]\}
\end{align}
Observe that indeed the roles of $0$ and $\infty$, of $N$ and $R$, as well as of $\min$ and $\max$ are exchanged.

As above, there are generalized transition Büchi conditions for every SCC.

\subsubsection{Output function}

To complete the description of how we ``implement'' the Main Theorem, we need to specify an appropriate output function $\lambda$. For every state $\{v_q\}_{p \in Q}$ we set 
\begin{align}
  \label{eq:40}
  \lambda(\{v_q\}_{q \in Q}) = \{q \in N \mid v_q < \infty\} \cup \{q \in R \mid v_q = \infty\} \enspace.
\end{align}
This is consistent with interpretation of the variables $v_q$ as explained in the previous subsections.

\subsubsection{Proof of correctness}
\label{sec:correctness}

First of all, it is easy to see that the automaton $\Aut B$ has the properties 1., 2., and 3. stated in the Main Theorem.

For the rest, it is enough to show the following for every word~$u$:
\begin{enumerate}[(i)]
\item If there is a final run $r$ of $\Aut B$ on $u$, then $\lambda(r(i)) = f _ {\Aut A}(u)$ for every $i \in \omega$.
\item There is a final run of $\Aut B$ on~$u$.
\item There is only one final run of $\Aut B$ on $u$.
\end{enumerate}

This is best proved by an induction on the SCC's of the transition graph of $\Aut A$, starting in the base case with the ``lowest'' SCC's, that is, the ones without outgoing edges.

We show how the inductive step goes for a non-recurring SCC; the base cases and the inductive step for recurring SCC's can be dealt with in a similar fashion, using duality where appropriate. In fact, the base cases are instances of the inductive step.

So in the following, $S$ is a non-recurring SCC. We refer to the SCC's reachable from $S$ as \emph{the other SCC's}, excluding $S$ itself, and denote the set of states of the other SCC's by~$T$. We assume that (i)--(iii) hold for the other SCC's (induction hypothesis).

\paragraph*{Proof of (i).}

Let $r$ be a final run of $\Aut B$ on~$u$ and write $\{v_{q,i}\}_{q \in Q}$ for $r(i)$. It is sufficient to show that for every $i$ and $q \in S$ the following hold.
\begin{itemize}
\item If $q \in N$ and $v_{q,i} < \infty$, then $u[i,\omega)$ is accepted from $q$.
\item If $q \in N$ and $v_{q,i} = \infty$, then $u[i,\omega)$ is not accepted from $q$.
\end{itemize}
We show how the proof goes for the case of the first item; the other case can be dealt with in a similar fashion, using duality. 

We construct a suitable part of a final run graph, which connects with the final run graph for the other SCC's, known to exist by induction hypothesis. To this end, we assign to each vertex of the graph to be constructed a value. For each variable $v_{q,i}$ with $q \in S$ we have the vertex $\langle  i, q\rangle$ and the value assigned to it is~$v_{q,i}$. The other vertices are intermediate vertices and constructed as follows, for each $v_{q,i}$ with $v_{q,i} < \infty$ and $q \in S$ separately, by an induction on $\delta(q)$. So the formulas $\tau$ dealt with in the following are all assumed to be subformulas of $\delta(q)$.

There are three base cases:
\begin{itemize}
\item If $\tau = B$ and $u(i) \in B$, then $\langle  i, B \rangle$ belongs to the run graph and is assigned $0$. 
\item If $\tau = \Next q'$ and $q' \in S$, then $\langle i+1, q' \rangle$ belongs to the run graph being constructed and has already been assigned $v_{q',i+1}$ (see above). No vertex is added.
\item If $\tau = \Next q'$ and $q' \notin S$, then $\langle i+1, q' \rangle$ belongs to the run graph known to exist by the induction hypothesis, provided 
  \begin{itemize}
  \item $q'$ is non-recurring and $v_{q',i+1}$ is finite or
  \item $q'$ is recurring and $v_{q',i+1}$ is infinite.
  \end{itemize}
  No vertex is added.
\end{itemize}
And there are two cases in the inductive step:
\begin{itemize}
\item If $\tau = \tau_0 \vee \tau_1$ and $\langle i, \tau_0 \rangle$ or $\langle i', \tau_1 \rangle$ is part of the run graph, then $\langle i, \tau \rangle$ is part of the run graph. It has an edge to the vertex with the smaller value assigned to it or to both vertices if they have the same value. The vertex $\langle  i, \tau \rangle$ is assigned the same value as its successor(s).
\item If $\tau = \tau_0 \wedge \tau_1$ and $\langle i, \tau_0 \rangle$ and $\langle i, \tau_1 \rangle$ are part of the run graph, then $\langle i, \tau \rangle$ is part of the run graph. It has an edge to both vertices and is assigned the maximum of the value of its successors.
\end{itemize}
Observe that this construction mimics the mechanics of the transition function. In particular, $\langle  i, \delta(q) \rangle$ is part of the run graph, the value assigned to this vertex is less than or equal to $v_{q,i}$, depending on the lifting, and the values along the paths do not increase. We can add the edges from $\langle i, q \rangle$ to $\langle  i, \delta(q) \rangle$, for each $q \in S$ and each $i$. 

Finally, observe that, by construction, if there is an infinite path through the part of the run graph just constructed, then one of the Büchi transition conditions of $\Aut B$ for the SCC $S$ is violated. This concludes the proof for the inductive step.

\paragraph*{Proof of (ii).}

The induction hypothesis is that we already have a run $r$ on the other SCC's. So when we write $v_{q,i}$ for $r(i)$, then every $v_{q,i}$ for $q \in T$ has already been defined. We need to extend $r$ to $S$ in the sense that we need to determine values $v_{q,i}$ for $q \in S$.

The key step is to define a monotone operation $F$ on families $\{v_{q,i}\}_{q \in S, i \in \omega}$ as follows. First, we write $\{v'_{q,i}\}_{q \in S, i \in \omega}$ for $F(\{v_{q,i}\}_{q \in S, i \in \omega})$. Second, we stipulate that $\{v'_{q,i}\}_{q \in S, i \in \omega}$ is obtained from $\{v_{q,i}\}_{q \in S \cup T, i \in \omega}$ by applying the transition function.

Clearly, the operation $F$ is monotone when we view the families $\{v_{q,i}\}_{q \in S, i \in \omega}$ as being ordered point-wise. This means that when we start from the family with $v_{q,i} = \infty$ for $q \in S$ and $i \in \omega$, then we eventually reach a fixed point. This fixed point satisfies the transition function (restricted to~$S$ and the other SCC's), just as any other fixed point does, but the least fixed point also satisfies the Büchi transition conditions for $S$. So it is a final run for $S$ and the other SCC's, which concludes the proof for the inductive step.

\paragraph*{Proof of (iii).}

Let $r$ be the run defined by the fixed point construction in the proof of (ii). By way of contradiction, let $r'$ be any other final run. We write $r(i)$  as $\{v_{q,i}\}_{q \in S \cup T}$ and $r'(i)$ as $\{v'_{q,i}\}_{q \in S \cup T}$. 

By induction hypothesis, the two runs can only differ for some $q \in S$. Since any run can be viewed as a fixed point (see the proof of (ii)) and $r$ is the least fixed point, there must be $q \in Q$ and $i \in \omega$ such that $v_{q,i} < v'_{q,i}$. Since the transition function is deterministic (and because of monotonicity), for every $j>i$ there must be some $q_j$ such that $v_{q_j,j} < v'_{q_j,j}$. Moreover, the corresponding pairs $\langle  j, q_j  \rangle$ can be assumed to be vertices of the graph constructed in the proof of (i). Because $q$ is assumed to be non-recurring, there is some $j'$ such that $q_{j'} \in S$ and 
\begin{itemize}
\item $\langle  j', q_{j'} \rangle$ is a dead end or
\item all successors of $\langle j', q_{j'} \rangle$ belong to $T$.
\end{itemize}
In both cases, the transition function would set $v_{q_{j'},j'}$ and $v'_{q_{j'}, j'}$ to the same value---the desired contradiction.

\section{Applications}

We present three applications of the Main Theorem in what follows. First, we use it to translate linear-time temporal formulas into backward deterministic $\omega$-automata. Second, we use it to translate alternation-free modal/temporal fixed-point formulas into backward deterministic $\omega$-automata. Third, we use it to transform a given Büchi automaton into an equivalent backward deterministic automaton.

\subsection{Linear-time temporal logic}

In this section, we explain that our construction is ``optimal'' with regard to converting linear-time temporal formulas into equivalent Büchi automata in the sense that first transforming a given formula of size $n$ into a weak alternating automaton and then applying our construction yields a backward deterministic generalized Büchi automaton with $2^n$ states, which is what one typically gets, see, for instance, \cite{vardi-wolper-1994}.

In the variant of linear-time temporal logic (LTL) we consider, formulas are built from letters over a given alphabet using boolean connectives and future temporal operators such as ``next'' ($\TLNext$), ``eventually'' ($\Eventually$), ``always'' ($\Always$), ``until'' ($\mathsf U$), and ``release'' ($\mathsf R$). As usual, we interpret such formulas in $\omega$-words over the given alphabet. All operators except ``next'' are interpreted non-strict, but what we describe works with minor modifications also for the strict variants.

Without loss of generality, we assume formulas are in negation normal form, that is, negation only occurs in front of letters of the alphabet. 

We next recall the transformation of LTL formulas into equivalent weak alternating automata, see, for instance, \cite{gastin-oddoux-2001}. A weak alternating automaton equivalent to a given formula~$\phi$ has a state~$q_\psi$ for every subformula $\psi$ of $\phi$; its transition function is defined by induction by the following equations:
\begin{align}
  \delta(a) & = \{a\}\\
  \delta(\neg a) & = A \setminus \{a\}\\
  \delta(q_{\psi \vee \psi'}) & = \delta(q_\psi) \vee \delta(q_{\psi'}) \\
  \delta(q_{\psi \wedge \psi'}) & = \delta(q_\psi) \wedge \delta(q_{\psi'})\\ 
  \delta(q_{\TLNext \psi}) & = \Next q_\psi\\
  \label{eq:20}
  \delta(q_{\Eventually \psi}) & = \delta(q_\psi) \vee \Next q_{\Eventually \psi} \\
  \label{eq:21}
  \delta(q_{\Always \psi}) & = \delta(q_\psi) \wedge \Next q_{\Always \psi}\\
  \label{eq:22}
  \delta(q_{\Until \psi {\psi'}}) & = \delta(q_{\psi'}) \vee (\delta(q_\psi) \wedge \Next q_{\Until \psi {\psi'}}) \\
  \label{eq:23}
  \delta(q_{\Release \psi {\psi'}}) & = \delta(q_{\psi'}) \wedge (\delta(q_\psi) \vee \Next q_{\Release \psi {\psi'}})
\end{align}
The states for eventually and until formulas are non-recurring; the states for always and release formulas are recurring. For the other states, it does not matter whether they are recurring or non-recurring, because they do not belong to any cycle in the transition graph.

Obviously, the automaton is a very weak alternating $\omega$-automaton. So the Main Theorem yields:

\begin{corollary}
  The transformation of an LTL formula of size $n$ into an equivalent backward deterministic generalized Büchi automaton via weak alternating automata results in an automaton with $2^n$ states.
\end{corollary}

\subsection{The alternation-free linear-time $\mu$-calculus}
\label{sec:mucalculus}

To begin with, we briefly describe the dialect of the linear-time $\mu$-calculus  \cite{vardi-1988, banieqbal-barringer-1987} we use. Given an alphabet $A$, the set of all linear-time $\mu$-calculus formulas (expressions), denoted $\NuTL$, is the smallest set consisting of
\begin{itemize}
\item $a$ and $\neg a$, for $a \in A$,
\item $X$, for $X \in \VarsTL$, where $\VarsTL$ is a supply of variables,
\item $\Next \phi$, if $\phi$ belongs to the set,
\item $\phi_0 \vee \phi_1$ and $\phi_0 \wedge \phi_1$, if $\phi_0$ and $\phi_1$ belong to the set,
\item $\mu_i \vec X . \vec \phi$ and $\nu_i \vec X. \vec \phi$, if $\vec X = \langle X_0, \dots, X_{r-1}\rangle$ is a vector of distinct variables from $\VarsTL$, $\phi_0, \dots, \phi_{r-1}$ belong to the set, and $i<r$.
\end{itemize}
In our vectorial fixed point dialect the formula $\sigma_i \vec X . \vec \phi$ refers to the $i$-th component of the least/greatest vectorial fixed point of $\vec \phi$.

Without loss of generality, we assume every variable is bound only once in every formula. As a consequence, every subformula $\sigma_i \vec X. \vec \psi$ of a given formula $\phi$ can then be referred to by $\phi_{X_i}$.

The vertex set of the \emph{dependence graph} of a formula $\phi$ is the set of subformulas of~$\phi$. Edges go
\begin{itemize}
\item from $\Next \psi$ to $\psi$,
\item from $\psi_0 \vee \psi_1$ and $\psi_0 \wedge \psi_1$ to $\psi_0$ and to $\psi_1$, and
\item from $\sigma_i \vec X . \vec \psi$ to $\psi_i$.
\end{itemize}
A formula $\phi$ has an \emph{alternation} \cite{niwinski-1986} if in its dependence graph there is a cycle with a $\mu$- and a $\nu$-subformula. We may assume that the formulas are such that the resulting automaton is \emph{guarded}, that is, we only consider guarded formulas \cite{banieqbal-barringer-1987}: in every cycle in the dependence graph there is a $\Next$-subformula.

Just as for LTL, there is a straightforward inductive translation from alternation-free closed $\NuTL$ formulas into weak alternating automata, where, again, for each subformula $\psi$ there is a corresponding state $q_\psi$, see, for instance, \cite{kupferman-vardi-wolper-2000,lange-2005}:
\begin{align}
  \delta(q_a) & = \{a\} \\
  \delta(q_{\neg a}) & = A \setminus \{a\}\\
  \delta(q_{\Next \psi}) & = \Next {q_\psi}\\
  \delta(q_{\psi \vee \psi'}) & = \delta(q_\psi) \vee \delta(q_{\psi'}) \\
  \delta(q_{\psi \wedge \psi'}) & = \delta(q_\psi) \wedge \delta(q_{\psi'})\\
  \delta(q_{\mu_i \vec X. \vec \psi}) & = \delta(q_{\nu_i X. \vec \psi}) = \delta(q_{\psi_i}) \\
  \delta(q_X) & = \delta(q_{\phi_X})
\end{align}
The states $q_\psi$ where $\psi$ is part of some cycle of a least fixed point formula in the dependence graph are non-recurring; the states $q_\psi$ where $\psi$ is part of some cycle with a greatest fixed point formula are recurring. For the other states, it does not matter whether they are recurring or non-recurring, because they do not belong to any cycle.

As every closed $\NuTL$-formula over some alphabet~$A$ is true or not in a position of a given $\omega$-word, a tuple $\vec \phi = \langle \phi_0, \dots, \phi_{k-1}\rangle$ of closed-$\NuTL$ formulas $\phi_0, \dots, \phi_{k-1}$ over~$A$ defines a function $f_{\vec \phi} \colon A^\omega \to (2^{\{0, \dots, k-1\}})^\omega$ by
\begin{align}
  f_{\vec \phi}(u)(i) = \{j < k \mid \text{$\phi_j$ is true for the suffix $u[i,\omega)$}\}
\end{align}
for every $u \in A^\omega$.

We obtain as an immediate consequence of the Main Theorem:
\begin{corollary}
  \label{thm:mu}
  For every tuple $\vec \phi$ of closed alternation-free $\NuTL$-formula $\phi_i$ there exists a backward deterministic automaton $\Aut B$ and an output function $\lambda$ for $\Aut B$ such that the function defined by $\vec \phi$ is the same as the function computed by $\Aut B$ with respect to $\lambda$, that is, $f _ {\vec \phi} = f _ {\Aut B, \lambda}$. The automaton $\Aut B$ has the following properties.
  \begin{enumerate}
  \item Let $S_0, \dots, S_{k-1}$ be an enumeration of all SCC's of the dependence graph of $\vec \phi$ and $m_i = S_i \cap Q$ for $i<k$. Then the number of states of $\Aut B$ is less than $\prod_{i<k} (m_i + 1)^{m_i}$, in particular, $(n+1)^n$ is an upper bound for the number of states of $\Aut B$ when $n$ is the number of states of $\Aut A$.
  \item The automaton $\Aut B$ has a generalized transition Büchi condition with as many Büchi sets as $\vec \phi$ has subformulas.
  \end{enumerate}
\end{corollary}

\subsection{From Büchi-automata to backward deterministic automata}

In this section, we show how our construction can be combined with rank functions to obtain a transformation from non-deterministic Büchi automata to equivalent backward deterministic automata. The main idea is to first describe the canonical rank functions by Kupferman and Vardi in the alternation-free linear-time $\mu$-calculus and then use Corollary~\ref{thm:mu}.

\subsubsection*{Background on canonical rank functions}

Let $\Aut A$ be a Büchi automaton given by some alphabet~$A$, a finite set $Q$ of states, a set of initial states $I \subseteq Q$, a transition relation $\Delta \subseteq Q \times A \times Q$, and a Büchi set $B \subseteq Q$. Further, let $u \in A^\omega$. 

We consider the run DAG of $\Aut A$ on $u$, which is the graph with vertex set $\omega \times Q$ and an edges as follows. For every $i$, the graph contains all edges from $\langle i, q\rangle$ to $\langle i+1, q'\rangle$ for $\langle q, u(i), q'\rangle \in \Delta$, and no other edges.

In general, a leveled DAG is a graph with vertex set $\subseteq \omega \times Q$ and where every edge is from some vertex $\langle i, q\rangle$ to some vertex $\langle i+1, q'\rangle$. Vertices in leveled DAG's are classified as follows. A vertex is called 
\begin{itemize}
\item \emph{finitary} if it has only a finite number of descendants,
\item \emph{$B$-tagged} if its second component belongs to $B$,
\item \emph{$B$-free} if none of its descendants (including itself) is $B$-tagged,
\item \emph{$B$-recurring} if there is an infinite path with infinitely many $B$-tagged vertices and starting in the vertex.
\end{itemize}
The \emph{ultimate width} of such a DAG is the limes inferior of the number of non-$B$-recurring infinitary vertices on a given level.

% Clearly, peeling a leveled DAG does not remove any $B$-recurring vertex, and if it does not change the DAG at all, then all vertices are $B$-recurring, because every vertex has a $B$-tagged strict descendant. Moreover, if there are non-$B$-recurring infinitary vertices, then peeling decreases the ultimate width by at least one, which leads to:

%, as explained in what follows.

Consider a non-$B$-recurring infinitary vertex. By König's lemma, there is an infinite path starting in it. Assume that every $B$-tagged strict descendant of the vertex is finitary. Then, after removing the finitary vertices, each successor of the vertex is $B$-free, but the infinite path is still there and all of its vertices (except, maybe, the first one) are removed in the second step, decreasing the ultimate width by one. If there is a $B$-tagged infinitary strict descendant of the vertex, apply the same argument to it. This cannot go ad infinitum, because a path with an infinite number of $B$-tagged vertices would be constructed.

So leveled DAG's is that they can be decomposed in a simple fashion by repeating the following operation, here called \emph{peeling}: first, remove all finitary vertices; second, remove all $B$-free vertices.

\begin{fact} \cite{kupferman-vardi-2001}
  For every Büchi automaton with $n$ states, peeling the run DAG of any $\omega$-word $n$ times yields the subgraph induced by the $B$-recurring vertices.
\end{fact}

By the above, each vertex $v$ in a run DAG can be assigned a value in $\omega \cup \{\infty\}$ according to when the vertex is removed by peeling the DAG successively. More precisely, when $i$ is a natural number and all vertices with value $< 2i$ are removed from the given DAG, the finitary vertices in the remaining DAG get assigned~$2i$; when all vertices with value $< 2i+1$ are removed, the $B$-free vertices in the remaining DAG get assigned~$2i+1$. The $B$-recurring vertices get assigned $\infty$. The number assigned to a vertex $v$ is called its \emph{canonical rank}, it is denoted $c(v)$, and, according to the above, it is $\infty$ or $< 2n$, when $n$ is the width of the DAG started with.

\begin{fact} \cite{carton-michel-2003}
  \label{fact:canonical}
  For a Büchi automaton with $n$ states, let $c$ be the canonical rank function of the run DAG of some $\omega$-word. The word is accepted from some state $q$ if, and only if, $c(\langle 0, q\rangle) = \infty$.
\end{fact}

\subsubsection*{Defining the canonical rank function in $\NuTL$} 

We next present $\NuTL$-formulas which define the canonical rank function for a given Büchi automaton $\Aut A$ and any $\omega$-word.

Assume $\Aut A$ has $n$ states, say $Q = \{q_0, \dots, q_{n-1}\}$. For every $i<2n$ and $j<n$, we define a formula $\chi_i^j$ such that if $\chi_i^j$ is interpreted in some $\omega$-word~$u$, then the value of $\chi_i^j$ is the set $\{k \in \omega \mid c(\langle k, q_j\rangle) \leq i\}$, where $c$ is the canonical rank function for the run DAG of $\Aut A$ on~$u$. In view of Fact~\ref{fact:canonical}, the tuple $\vec \chi$ defined by $\vec \chi = \langle \chi_{2n-1}^0, \dots, \chi_{2n-1}^{n-1}\rangle$ defines the function we are interested in, more precisely, when each $\chi_{2n-1}^j$ is replaced by a formula denoting the negation of $\chi_{2n-1}^j$, which can easily be achieved.

We set 
\begin{align}
  \chi_i^j = \sigma_i \langle X_i ^ 0, \dots, X_i^{n-1}\rangle . \langle \phi_i^0, \dots, \phi_i^{n-1} \rangle \enspace,
\end{align}
where the $X_i^j$'s are distinct variables and the $\sigma_i$'s and $\phi_i^j$'s are specified in what follows, by induction on~$i$.

The formulas of the base case, $i=0$, are defined by
\begin{align}
  \phi_0^j = \bigvee_{a \in A} \left(a \wedge \bigwedge_{\langle q_j, a, q_k\rangle \in \Delta} \Next X_0 ^ k\right) \enspace.
\end{align}
The formulas essentially say that every vertex without successors has rank~$0$, and every vertex whose successors have rank~$0$ does so, too.

In the inductive step, we use $\vec {X_i}$ to denote the tuple $\langle X_i^0, \dots, X_i^{n-1}\rangle$.
We distinguish between $i$ odd and $i$ even. For odd~$i$, say $i = 2i'+1$, we set $\sigma_{2i'+1} = \nu$ and
\begin{align}
  \phi_ {2i'+1} ^ j & = 
  \begin{cases}
    \chi _ {2i'} ^ j \enspace, & \text{if $q_j \in B$,}\\
    \displaystyle \chi _ {2i'} ^ j \vee \bigvee _ {a \in A} \left(a \wedge \bigwedge_{\langle q_j, a, q_k\rangle \in \Delta} \Next X _ {2i+1} ^
      j \right)\enspace, & \text{otherwise ($q_j \notin B$).}
  \end{cases}
\end{align}
So the $\chi_{2i'+1}^j$'s are to be read as follows: a $B$-tagged vertex has rank $\leq 2i'+1$ only if its rank is $\leq 2i'$; a non-$B$-tagged vertex has rank $\leq 2i'+1$ if it has rank $\leq 2i'$ or all its descendants have rank $\leq 2i'+1$. 

For even~$i$, say $i = 2i'$ and $i' > 0$, we set $\sigma_{2i'} = \mu$ and
\begin{align}
  \phi_{2i'}^j = \chi _ {2i'-1} ^ j \vee \bigvee_{a \in A} \left(a \wedge \bigwedge_{\langle q_j, a, q_k\rangle \in \Delta} \Next X _ {2i'} ^ k\right) \enspace,
\end{align}
which is very similar to the formulas to the formulas in the base case.

A straightforward application of Corollary~\ref{thm:mu} to the above formulas leads to an upper bound of $O(n^2)^{O(n^2)}$, because $\vec \chi$ is of size quadratic in~$n$ and has $2n^2$ variables. Observe, however, the following.
\begin{itemize}
\item Assume a fixed point formula is such that its dependence graph has the property that every path from a next subformula to another next subformula passes through a fixed point subformula. Then an optimized transformation into a weak alternating automaton results in as many states as there are fixed point variables. That is, we obtain only $2n^2$ states for the automaton corresponding to $\vec \chi$.
\item The size of each SCC in such an automaton for $\vec \chi$ is~$n$, because we have $n$ variables in every vectorial fixed point subformula.
\item For a fixed $j$, the values of the fixed point expressions $\chi_i^j$ are pairwise disjoint sets.
\end{itemize}

This all implies:

\begin{corollary}
  The transformation of a non-deterministic Büchi automaton with $n$ states into a backward deterministic generalized transition Büchi automaton via $\NuTL$ and weak alternating automata results in a backward deterministic generalized transition Büchi automaton with $O((2n(n+1))^n)$ states.
\end{corollary}

\section{Conclusion}

The translation from weak alternating to backward deterministic automata presented in the Main Theorem is indeed a general construction in the sense that from it other translations into backward deterministic $\omega$-automata can be derived. It remains open whether the Main Theorem and the translation presented in the section on applications can be fine-tuned (or improved) in such a way that the best known bounds can be met.

\bibliographystyle{plain}
\bibliography{linearMu}

\end{document}